\documentclass[twocolumn]{aastex701}

\begin{document}

\title{Magnetic Flux Tubes Illuminated by Pulsar Winds}

\author[orcid=0000-0001-5874-1432]{Yifan Sun}
\affiliation{Department of Physics, The University of Hong Kong, Pokfulam Road, Hong Kong}
\affiliation{Hong Kong Institute for Astronomy and Astrophysics, The University of Hong Kong, Pokfulam Road, Hong Kong}
\email[show]{u3009913@connect.hku.hk}

\author[orcid=0000-0002-5847-2612]{C.-Y. Ng}
\affiliation{Department of Physics, The University of Hong Kong, Pokfulam Road, Hong Kong}
\affiliation{Hong Kong Institute for Astronomy and Astrophysics, The University of Hong Kong, Pokfulam Road, Hong Kong}
\email{ncy@astro.physics.hku.hk}

\author[orcid=0000-0003-1039-9521]{Siming Liu}
\affiliation{School of Physical Science and Technology, Southwest Jiaotong University, Chengdu, 610031, People’s Republic of China}
\email{liusm@swjtu.edu.cn}

\begin{abstract}
Observations of linear structure connecting  pulsars to gamma-ray halos
reveal injection of TeV electrons into the interstellar medium (ISM). In some cases, 
this could be attributed to nearly scattering-free electron transport along large-scale magnetic fields connected to pulsar winds with very slow diffusion across the field lines. In this work we model this process with a magnetic flux tube emerging from the pulsar and
attached to the ISM magnetic field. We show that particles in this case have
an anisotropic distribution of magnetic pitch angle, such that the overall emission
is highly beamed.
We apply this model to pulsar tails and filaments, including the extended X-ray
and TeV emission associated with PSR J1740+1000 and the misaligned X-ray jet in
the Guitar Nebula, to constrain their particle population and magnetic fields.
\end{abstract}


\section{Introduction}

Evolved pulsar wind nebulae (PWNe) that have left the parent supernova remnant
often show bow-shock structure and a long tail, since they travel
supersonically through the interstellar medium (ISM). These tails trace
the pulsar motion over tens of parsecs and consist of synchrotron-emitting
particles that can be observed in radio and/or X-ray. Recent numerical
simulations suggest that the magnetic field should align with the tail, 
as it is formed by the swept-up ISM magnetic fields left behind \citep{2019MNRAS.488.5690O}.
A few bow-shock PWNe are also found to be associated with diffuse gamma-ray emission
along the tail direction \citep[e.g.,][]{2011A&A...528A.143H,2025Innov...600802.}.
Spectral fitting suggests that the pulsar is able to provide enough high-energy
particles to produce the observed gamma-ray emission via inverse Compton (IC)
scattering. This however requires the particles to keep most of their energy during
transport through the tail. To explain the multiwavelength morphology
of these TeV sources, there have been previous attempts to consider complicated
magnetic field geometry with a significantly aligned component to
transport the particles from the PWN to TeV-emitting regions
\citep[e.g.,][]{2025PhRvL.135s1002B,2025ApJ...987...19Y}.

Another related structure is misaligned X-ray filaments in bow-shock PWNe.
\citep[see][]{2024ApJ...976....4D}. They are highly collimated, much narrower than
pulsar tails, and extend to a few parsecs. More intriguingly, they often show 
large misalignment angle with the pulsar proper motion direction, and some
even point ahead of the pulsar. They are only detected in X-rays with
no radio counterparts, and their spectra are well described by a single power law
without cooling. Such linear structure is suggested to be due to high-energy
particles injected into pre-existing magnetic field lines in the ISM when
the PWN passes by \citep{2008A&A...490L...3B,2019MNRAS.490.3608O,2024A&A...684L...1O}.
Previous studies modeled these with aligned magnetic fields \citep[e.g., the ``magnetic
bottles/antibottles'';][]{2019MNRAS.485.2041B} with a coherence length much larger
than the physical length size of the filaments, as indicated by the minimal
cooling observed.

In this paper, we model these linear structures using magnetic flux tubes. We
consider a one-zone leptonic model, where the same populations of electrons 
produce both the synchrotron (radio to X-ray) and IC (gamma-ray) emission. The
magnetic field is assumed to be well aligned and the particles are bound to the
field lines, following a near ballistic trajectory. We demonstrate that this
setup can produce an electron population with anisotropic pitch angle
distribution, and the emissions are beamed along the magnetic field lines.
This modifies the synchrotron emissivity and creates unique observational
signatures. One important consequence is that part of the flux tube
could be unobservable due to the beaming and the view angle. This can result in
a hard cutoff in the observed image. There have been studies of
anisotropic leptonic emission in the context of other astrophysical sources,
e.g., in pulsar emission \citep{1973ApJ...183..593E,1973ApJ...183..611E},
blazers, and gamma-ray bursts \citep{2018ApJ...864L..16Y}. These sources are more
beamed such that the anisotropy is more apparent. In previous works on PWNe,
however, this anisotropy is in general not considered, and the electrons are
often assumed to be isotropic. The model by itself is extremely versatile. In
addition to the abovementioned two types of sources, it can in principle be applied
to any source where the magnetic field lines are sufficiently aligned, i.e.,
when the physical length scale is much smaller than that of the magnetic field
turbulence.

In Section \ref{sec:model}, we discuss the basic setup and observational signatures of the magnetic flux tube model. We apply the model to two different type of sources: in Section \ref{sec:j1740}, we study the PWN tail of PSR J1740+1000 which is found to be associated with an offset TeV gamma-ray source, and in Section \ref{sec:guitar}, we study the misaligned filament formed by outflow from bow-shock PWNe, with Guitar Nebula as an example.

\section{Magnetic flux tube Model}\label{sec:model}

\subsection{Assumptions and setup}

\begin{figure}
\includegraphics[width=\columnwidth]{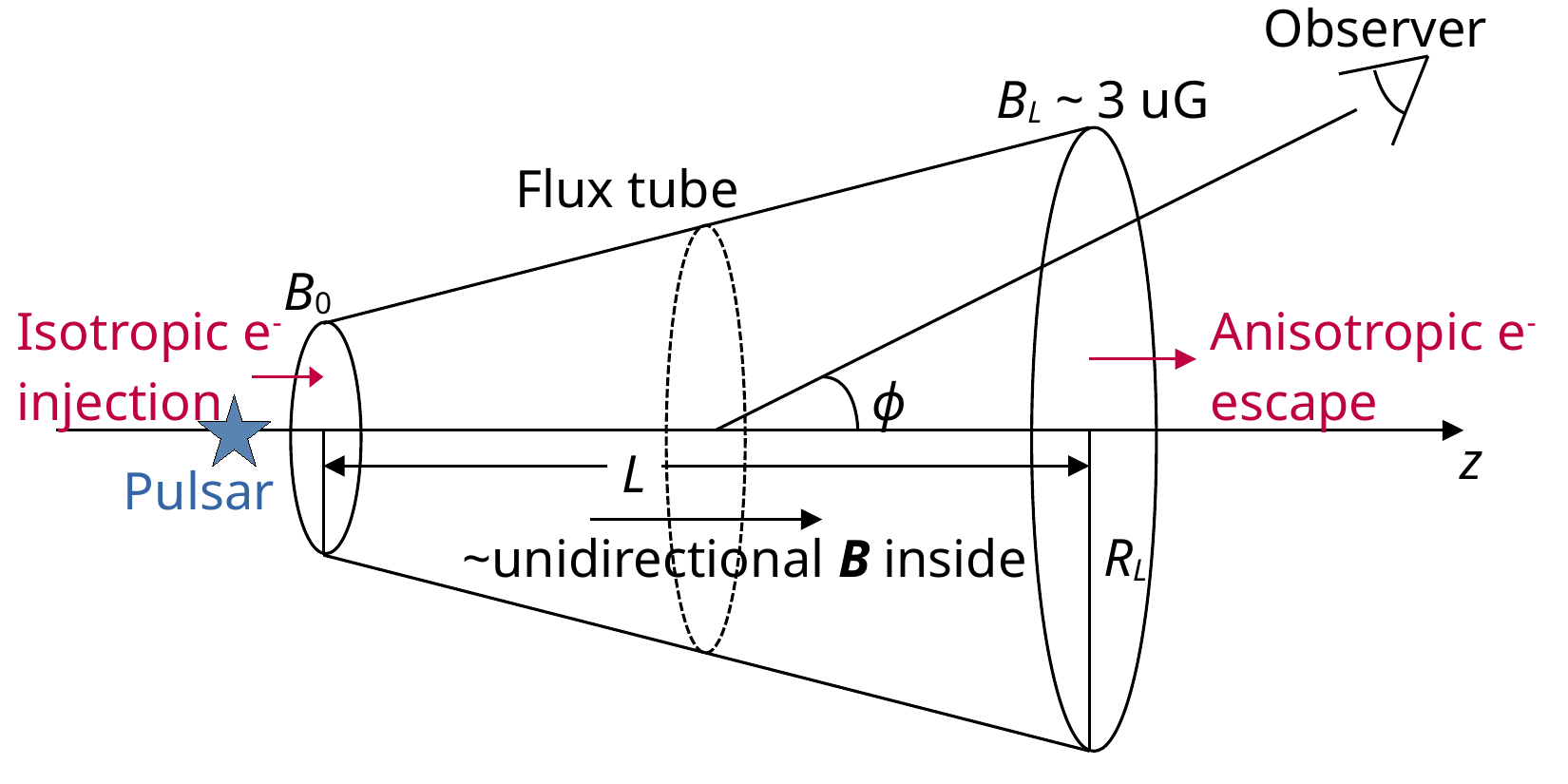}
\caption{Illustration of magnetic flux tube model. High-energy electrons
accelerated in the PWN are injected into the magnetic flux tube and transport to
the right. The magnetic field is assumed to be pointing to $+z$-direction 
and $B\propto z^{-2}$ for this figure. The flux tube has length $L$ and radius
$R_L$ at the exit end. $\phi$ is the global viewing angle defined using the axis.
\label{fig:flux_tube_illustration}}
\end{figure}

The geometry of the magnetic flux tube is
shown in Figure~\ref{fig:flux_tube_illustration}. High-energy particles are
injected from the pulsar end of the tube, where the magnetic field strength is
larger than the exiting end. The latter is assumed to have a typical ISM field
strength of a few $\mu$G. The $B$-field inside the tube is taken to be nearly
unidirectional, with slight divergence due to the decreasing field strength and
conservation of magnetic flux. In the figure we illustrate the case that the
field strength scales as the distance $z$ from the pulsar.\ $B\propto z^{-2}$.
This gives a flux tube with straight edges, but we note that this dependence can
be arbitrary in general. The particles inside the tube are bounded to the
magnetic field lines without any scattering. This means that the energetic
relativistic particles effectively move across the tube at nearly speed of
light. The physical size of a magnetic flux tube is limited by the ISM
turbulence length scale, which is suggested to be up to several hundred parsecs
\citep{2016JCAP...05..056B}. The particle traversal time is then $\sim$1\,kyr,
significantly shorter than the typical synchrotron cooling times of $\tau_{\rm
syn}=13.0(E_e/40\,{\rm TeV})^{-1}(B/5\,{\rm \mu G})^{-2}$\,kyr. We therefore
ignore any radiative energy loss in the flux tube and consider a
time-independent steady-state solution. The viewing angle $\phi$ between the
line of sight and the symmetry axis will significantly alter the observed source
morphology and other physical parameters.

\subsection{Particle evolution}
\label{s_evo}
The evolution of particle pitch angle distribution in a magnetic field is described by a simplified Vlasov equation \citep{2006JGRA..111.8101Q}:
\begin{equation}
    \mu v\frac{\partial f}{\partial z}+\frac{d\mu}{d t}\frac{\partial f}{\partial\mu}=0,
\end{equation}
where $f(z,\mu)$ is the distribution function of particles and $\mu=\cos{\alpha}$ is the cosine of
the local magnetic pitch angle $\alpha$.

The
term $d\mu/dt$ describes the pitch angle evolution. For a slowly varying
magnetic field, it can be shown that
the magnetic moment of a particle, which is proportional to $v_\perp^2/B$, is
conserved \citep{2011hea..book.....L}. Taking the time derivative
gives the following expression due to the so-called magnetic mirroring effect,
\begin{equation}
    \frac{d\mu}{dt}=-\frac{1-\mu^2}{2B}\frac{\partial B}{\partial z}.
\end{equation}
Note that the spatial variation of the magnetic field $B(z)$ will be chosen later
depending on the source geometry. The above two equations can be combined to form a
single partial differential equation, which can be solved to obtain
\begin{equation}
    \label{eqn:pde_sol}
    f(z,\mu)=f_0\left(\sqrt{1+\frac{\mu^2-1}{a}}\right)
\end{equation}
using the initial condition $f(0,\mu)=f_0(\mu)$ at the boundary $z=0$, where $a\equiv B(z)/B(0)$.
This suggests that the particle distribution can also be viewed as a function of the
magnetic field strength. In the following discussion, we assume $B$ is a decreasing function of $z$ and hence $a<1$.

One particularly simple but physically interesting model is isotropic particle injection at $z=0$. 
In this case, $f(0,\mu)$ is a constant independent of $\mu$. 
Since $\mu\in\left[\sqrt{1-a},1\right]$ in Equation~\ref{eqn:pde_sol}, 
the pitch angle distribution function $f(z,\mu)$ will be constrained within a
cone of $\sin{\alpha}\leq\sqrt{a}$. This cutoff value is a function of the
magnetic field strength and hence depends on the physical distance traveled.
There will be no particles with pitch angle larger than this cutoff value.
Inside this cone, the number of particles with pitch angle
$[\alpha,\alpha+d\alpha]$ remains constant. This shows that, in this simple
model of a flux tube with a decreasing magnetic field, the particles tend to
form a beam along the magnetic field line with a decreasing cutoff of the pitch
angle.

\subsection{Emission from anisotropic electron population}

The anisotropic pitch angle distribution of electrons inside the magnetic flux
tube can change the emissivity of both the synchrotron and IC emission from the standard
leptonic model. For the former,
the radiation power per unit solid angle
$d\Omega$ per unit frequency $d\omega$ for a single electron of energy $\gamma
m_e c^2$ and pitch angle $\alpha$ is given by
\citep[e.g.,][]{1979rpa..book.....R,2018ApJ...864L..16Y},
\begin{eqnarray}
    \label{eqn:syn}
    \frac{dW}{d\omega d\Omega}=\frac{\omega_B}{2\pi\sin^2{\alpha}}\frac{e^2}{3\pi^2c}\left(\frac{\omega\rho}{c}\right)^2\left(\frac{1}{\gamma^2}+\theta^2\right)^2 \nonumber\\
    \left[K^2_{2/3}(\xi)+\frac{\theta^2}{(1/\gamma^2)+\theta^2}K^2_{1/3}(\xi)\right].
\end{eqnarray}
In this equation, $\omega_B$ is the gyro frequency, $\rho$ is the radius of
curvature of the electron, $\theta$ is the angle between line of sight and the
electron's velocity vector, and $K_\nu(\xi)$ is the modified Bessel function
with the argument $\xi=(\omega\rho/3c)(1/\gamma^2+\theta^2)^{3/2}$. As
mentioned, the electron population will have a pitch angle distribution of
$\sin{\alpha}\leq\sqrt{a}$. If we assume an exponential cutoff powerlaw
distribution of electrons for PWNe, with $dN/dE\propto
E^{-1}\exp(-E/110~{\rm TeV})$, Equation~\ref{eqn:syn} can be integrated over
the electron's pitch angle $\alpha$ and energy $\gamma$ to obtain the emission
coefficient.

\begin{figure}
\includegraphics[width=\columnwidth]{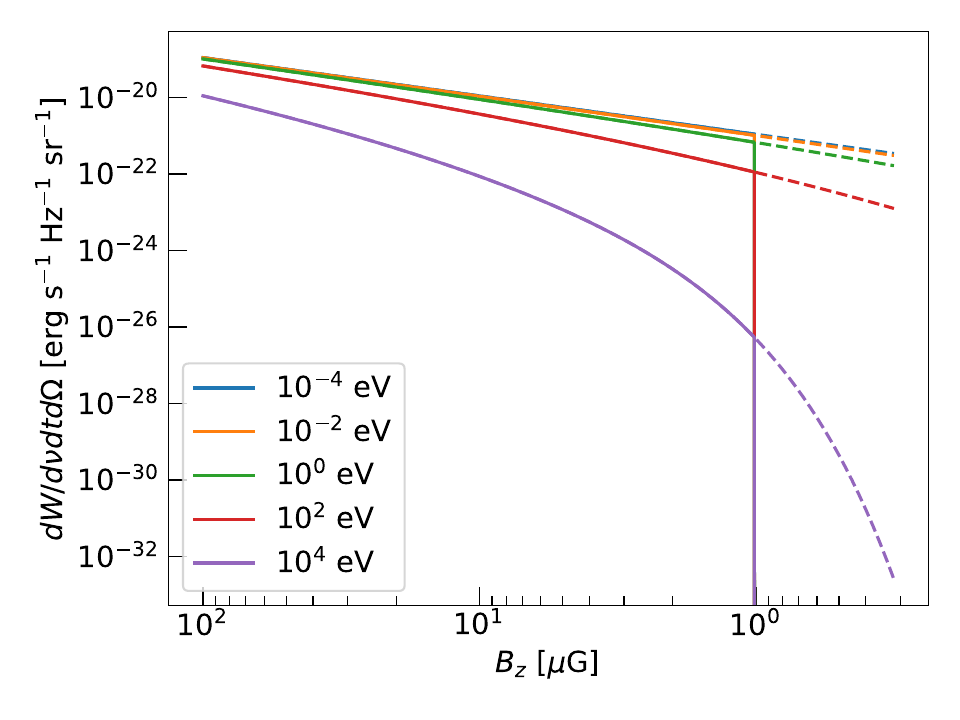}
\caption{Synchrotron spectral intensity, i.e., energy per unit time per
steradian per frequency, for a group of particles evolving along the magnetic
flux tube at different photon energies. This assumes a view angle of $
\phi=0.1$ and the magnetic field strength of $B_0=100\,{\rm\mu G}$ at the injection
end. The electron distribution is taken as an exponential cutoff power law 
$dN/dE\propto E^{-1}\exp(-E/110\,{\rm TeV})$. The solid lines show the emission
from the predicted evolution of an isotropic electron population injection. The
dashed line shows the emission from the same population of electrons but without
pitch angle evolution, consistent with standard semianalytical formula. They
agree until some cutoff point, where the pitch angle of all particles becomes
smaller than the view angle $\phi$, indicating a finite viewable length of the
magnetic flux tube.
\label{fig:syn_pitch_angle_dep}}
\end{figure}

Figure~\ref{fig:syn_pitch_angle_dep} shows the results for $\phi=0.1$ and
$B_0=100$\,$\mu$G. We also plot the emission from an electron population where
such effect is ignored. The two are normalized such that the number of particles
inside the cone $\sin{\alpha}\leq\sqrt{a}$ are the same. The plot shows
a hard cutoff in the observed emission when $\sin{\phi}<\sqrt{a}$,
i.e., all electrons evolve to have pitch angles smaller than the view angle.
This is due to beaming of the synchrotron emission, such that the
emission is confined within a cone of $\sim 1/\gamma$ along the particle
velocity direction, where $\gamma$ is the particle's Lorentz factor. Because of
the cutoff, the physical length of the magnetic flux tube can be longer than
what is observed. We also note that if the flux tube is pointing away from us,
i.e., $\phi>\pi/2$, all the emission will not be observable unless some other
effects, such as scattering, that randomize the pitch angle distribution.

For the IC process with cosmic microwave background photons, while it is
independent of the magnetic field and the pitch angle, the emission is still
expected to be beamed along the particle velocity direction due to anisotropic particle momentum distribution.
This modifies the IC emission profile 
\citep[e.g.,][]{2023PhRvD.107f3026L}. Trying to solve this numerically for arbitrary
electron and seed photon distributions proved to be computationally infeasible
for our application, due to the number of parameters that needed to be
integrated over. We note that the upscattered photons from these electrons are
beamed within an angle of $\sim1/\gamma$, given by the Lorentz transformation
between the electron rest frame and the observer frame
\citep{1979rpa..book.....R}. This allows us to approximate the IC emission to be
along the electron's velocity vector, and the directional dependent emission
profile is directly proportional to the number of particles moving along the line of
sight. There will be a similar cutoff as in the synchrotron process.

\subsection{Observational signatures}

\begin{figure*}
\gridline{
\fig{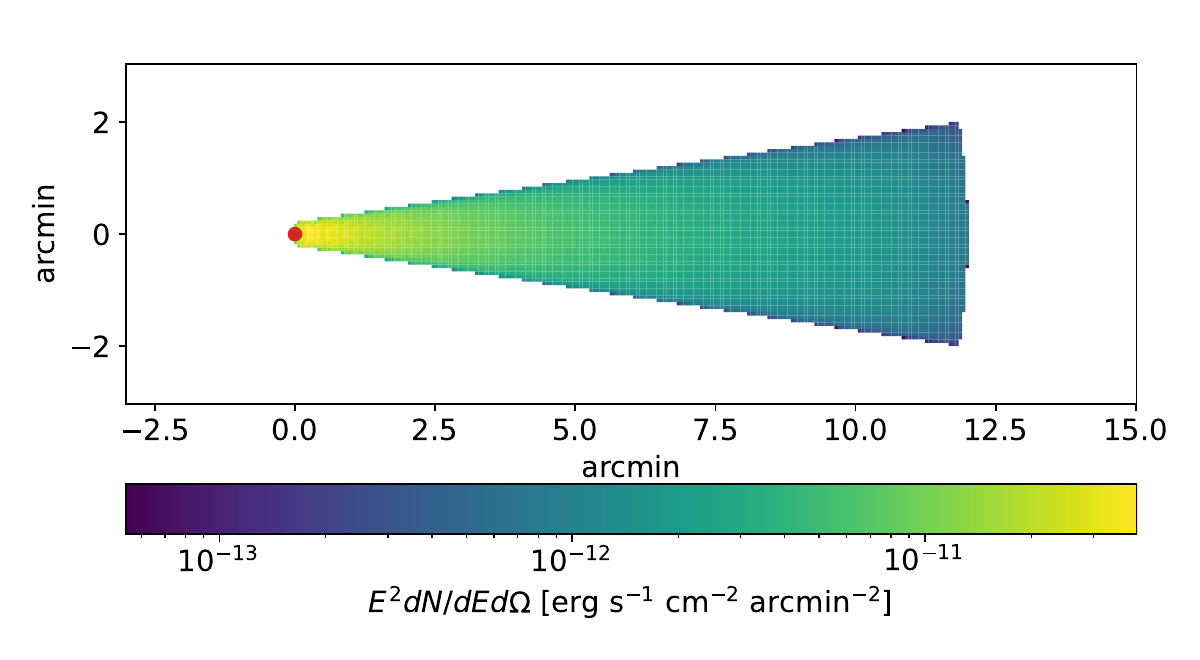}{0.48\textwidth}{(a) $\phi=\pi/30$, synchrotron at 1\,keV}
\fig{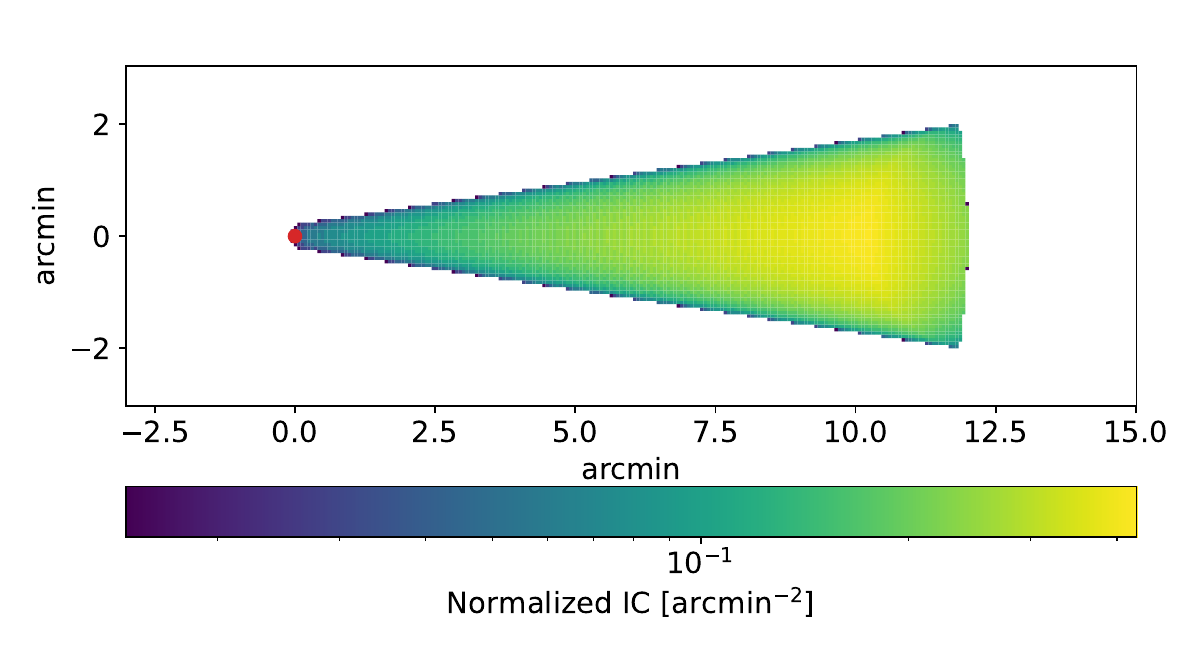}{0.48\textwidth}{(b) $\phi=\pi/30$, IC}
}
\gridline{
\fig{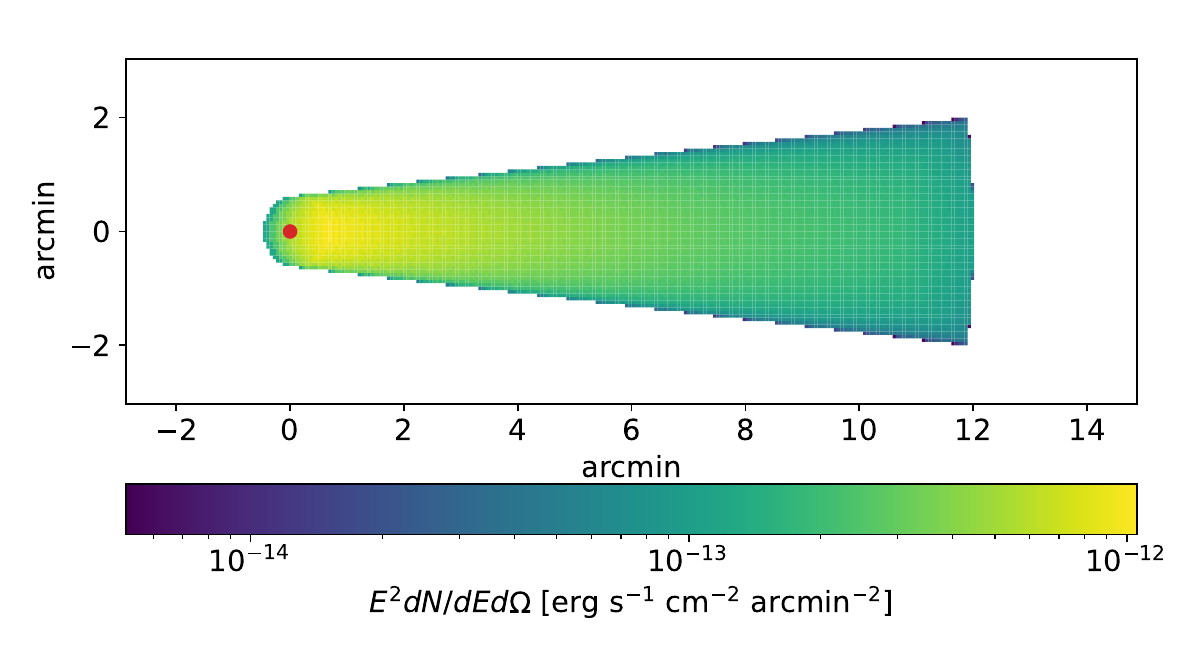}{0.48\textwidth}{(c) $\phi=\pi/10$, synchrotron at 1\,keV}
\fig{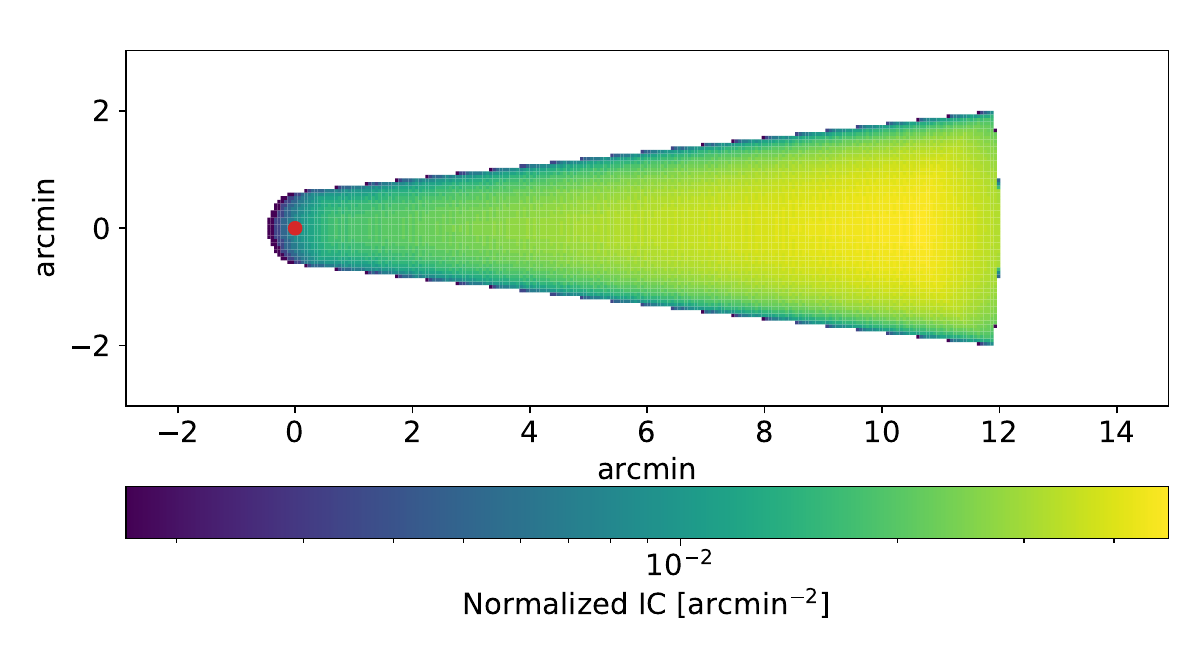}{0.48\textwidth}{(d) $\phi=\pi/10$, IC}
}
\gridline{
\fig{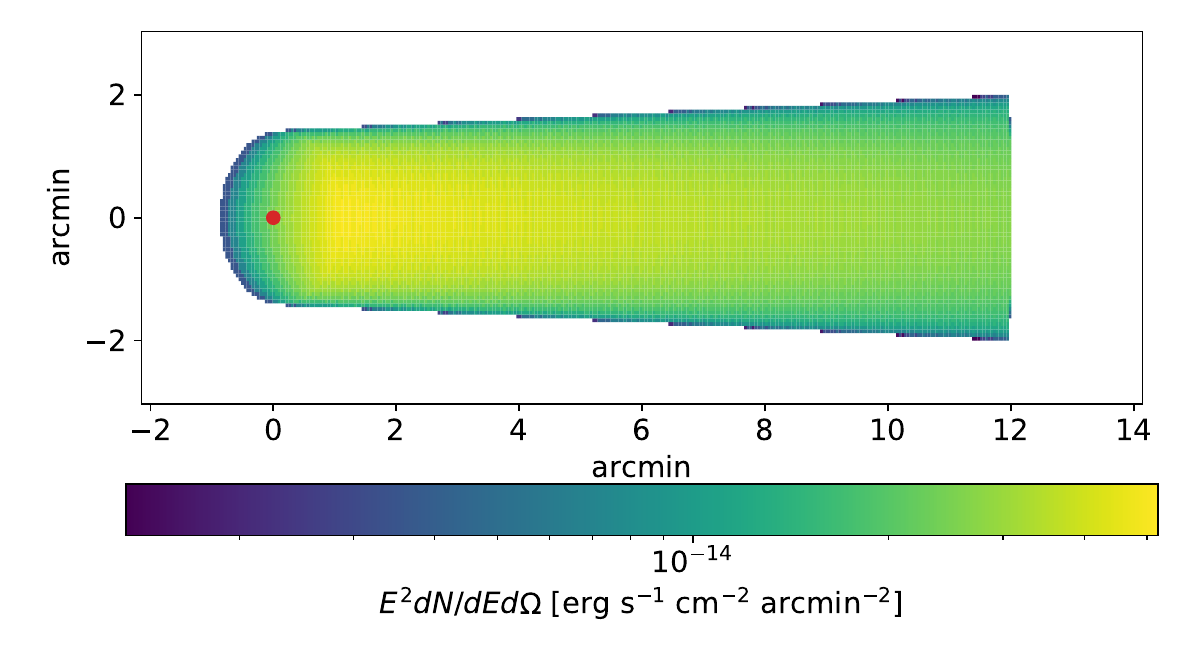}{0.48\textwidth}{(e) $\phi=\pi/4$, synchrotron at 1\,keV}
\fig{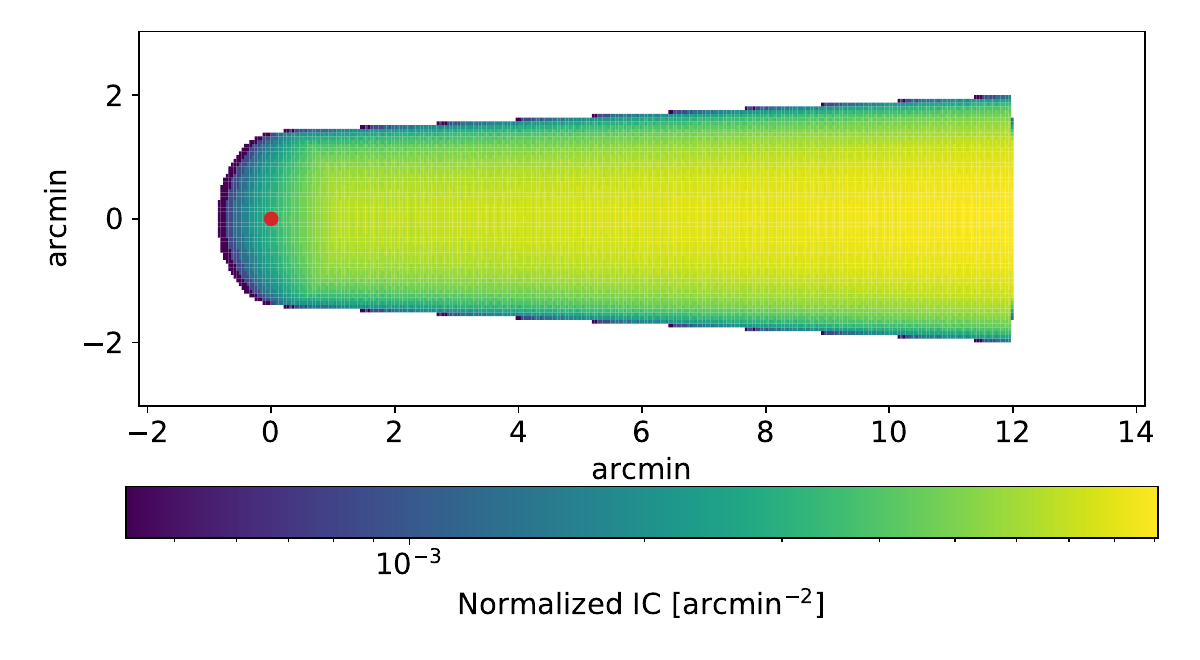}{0.48\textwidth}{(f) $\phi=\pi/4$, IC}
}
\caption{Simulated synchrotron X-ray images at 1\,keV (left column) and IC
gamma-ray images (right column) for the magnetic flux tube viewed at different
$\phi$. The red dots indicate the center of the injection end. We assume
$R_L=2'$, $B_L=3\,{\rm \mu G}$, and an electron distribution of $dN/dE\propto
E^{-1}\exp(-E/110\,{\rm TeV})$. The projected length is fixed at $12'$ by
changing the physical length of the flux tube with $L=12'/\sin{\phi}$. The
magnetic field at the injection end is set by $B_0=B_L/\sin^2{\phi}$.
\label{fig:flux_tube_image}}
\end{figure*}

To compare with observations, we adopt $B\propto z^{-2}$ which gives a truncated
cone shape flux tube as shown in Fig.~\ref{fig:flux_tube_illustration}. We
divide the tube into small discrete volumes and calculate the particle pitch
angle distribution. The synchrotron emission in each cell is computed using the \texttt{naima}
package \citep{2010PhRvD..82d3002A,2015ICRC...34..922Z} and then all cells
are summed over the line of sight to obtain a sky image.
Figure~\ref{fig:flux_tube_image} shows the resulting images with different viewing angles.
It is clear that the X-ray emission always peaks near the injection end, while
the IC emission peaks near the exit end. The former is due to the stronger
magnetic field strength, and the latter can be attributed to combination of
projection effect and the decreasing pitch angle of the particles, as discussed
in Section~\ref{s_evo}.

In the following discussion, we take the cutoff point as the exit end of the
flux tube, since the rest is not observable. The viewing angle $\phi$ is an
important parameter in the model that sets the $B$-field ratio between the
injection and exit ends as $B_0=B_L/\sin^2\phi$. For typical ISM field strength
at the exit end, e.g., 3\,$\mu$G in the case of Figure
\ref{fig:flux_tube_image}, this requires $B\sim$10--100\,$\mu$G at the injection
end for moderate $\phi$. Due to conservation of magnetic flux, the equation
above also determines the radius ratio between the two ends, $R_0/R_L$, and
hence the emission morphology. A small $\phi$ gives $R_0/R_L\ll 1$, resulting in
a short, cone-like X-ray tail with large opening angle (see
Figures~\ref{fig:flux_tube_image}a and b). On the other hand, if viewed edge-on,
i.e., large $\phi$, the X-ray emission would have constant width with a more
gradual decrease in the surface brightness (see
Figures~\ref{fig:flux_tube_image}e and \ref{fig:flux_tube_image}f).

As argued above, the effect of radiative cooling should be minimal during the
short traversal time of the particles in the flux tube. We therefore do not
expect significant variation of the emission spectrum. However, the reducing
$B$-field would shift the high-energy cutoff to lower energies. Such effect
could be observable with spatially resolved spectrum obtained from deep X-ray
observations. Moreover, the ordered magnetic field in the flux tube would
produce synchrotron emission with linear polarization fraction close to the
maximum allowed value \citep[see][for details]{2025MNRAS.542..902L}.

Finally, we note that in the simulated images we did not consider the effect of
finite gyroradius for the synchrotron-emitting particles. Accounting for this
will smooth out the edge of the flux tube in the image. For electrons emitting
photons of energy $E_{\rm keV}$, the gyroradius has an angular scale of
\begin{equation}
    \theta=26''E_{\rm keV}^{1/2}B_{\rm\mu G}^{-3/2}d_{\rm kpc}^{-1}
\end{equation}
\citep{2022ApJ...939...70D}.
This is a few arcseconds for $B\sim10$\,$\mu$G and $d\sim1\,$kpc, comparable with
the resolution limit of X-ray telescopes such as the Chandra X-ray Observatory
(CXO). In the sources that we will discuss in the following sections, such smoothing effect
can be safely ignored.

\section{Application 1: PSR J1740+1000 tail}\label{sec:j1740}

\subsection{Multiwavelength observations of J1740}

PSR J1740+1000 (hereafter J1704) is a middle-aged pulsar with characteristic age
$\tau_c=114\,{\rm kyr}$, spin-down luminosity $\dot{E}=2.3\times10^{35}\,{\rm
erg~s^{-1}}$ and dispersion measure-inferred distance of 1.2\,kpc 
\citep[][and references therein]{2021ApJ...916..117B,2022MNRAS.513.3113R}.
It has been observed in X-rays with CXO and XMM-Newton. As shown in
Figure~\ref{fig:j1740}, the XMM image reveals a $5.5'$ long tail
trailing the pulsar \citep{2008ApJ...684..542K}. The high Galactic latitude
($b=20.268^\circ$) of J1740 indicates that it is likely moving at a high speed
out of the Galactic plane, which can be confirmed with future proper motion
measurement. The extended X-ray emission is therefore suggested
to be a bow-shock PWN \citep{2021ApJ...916..117B}. The tail shows a nearly conical
shape at the tip with a width of $\sim1.5'$, then beyond 3\arcmin\ from the pulsar the width stays mostly
constant, and the orientation slightly bends to the west. The XMM data found a
power-law spectrum with $\Gamma=1.75\pm0.03$ for the overall $5.5\arcmin$ tail
\citep{2021ApJ...916..117B}. Recently, a reanalysis of the same XMM data shows
no significant spectral variation along the tail \citep{2025A&A...704A..30B},
suggesting minimal cooling for the particles. The CXO observation of the tail,
on the other hand, suggests a harder spectrum with $\Gamma=1.2\pm0.2$ in a
shorter length of $2.6\arcmin$ \citep{2024ApJ...976....4D}. In addition, we
analyze radio observation of the tail taken with the Karl G. Jansky Very Large Array (VLA) at
the 10\,cm band. The detailed analysis and results are described in the
Appendix~\ref{sec:vla}, and the total intensity map is shown in
Figure~\ref{fig:j1740}. No radio emission is found from the tail and we obtained
a 3$\sigma$ upper limit of $3.5$\,mJy.

Recently, TeV gamma rays are detected with LHAASO
near J1740. The emission peaks at $12'$ from the pulsar along the X-ray tail direction
and is slightly extended with angular scale $\lesssim 0.147^\circ$
\citep{2025Innov...600802.}. The spatial distribution of the TeV emission can be described using a diffusion-dominated model. The corresponding diffusion coefficient for energy $E>25\,{\rm TeV}$ is $D=3.55\times10^{26}\,{\rm cm^2~s^{-1}}$. A phenomenological fitting of TeV spectrum using a simple one-zone leptonic model gives an electron distribution of $dN/dE\propto E^{-p}\exp(-E/E_{\mathrm{cut}})$, with the index $p=2$ and a cutoff energy of $E_{\mathrm{cut}}=110$\,TeV. The nondetection of X-ray counterpart implies a low magnetic field
strength $\lesssim1.4\,{\rm\mu G}$ at the TeV center
 \citep{2025A&A...704A..30B}. 

\subsection{Model setup with an additional diffusion zone}

Assuming that the observed X-ray and gamma-ray emission trailing J1740 is powered
by the same population of particles injected by the PWN, we employ the
magnetic flux tube model to explain the offset between the two emissions. Our
model considers only the ballistic transport of highest energy particles but ignores
advection of particles strongly coupled with the background plasma as suggested by magnetohydrodynamics (MHD) simulations of bow-shock PWNe
\citep[e.g.,][]{2019MNRAS.484.4760B}. While advection could be the dominant
effect for low-energy electrons, we argue that it cannot solely explain the
transport of the $>10$\,TeV electrons that are responsible for the offset gamma-ray
emission. This initial postshock flow speed can be very high ($\approx c$) near
the pulsar, but it decelerates downstream to nearly the pulsar speed (several
hundreds km\,s$^{-1}$) as both numerical simulations and observations show
\citep{2010ApJ...712..596N,2015MNRAS.454.3886M}. As a result, the traversal
timescale over the parsec-long tail would be much longer than the cooling
timescale of the $>10$\,TeV electrons, hence requiring an alternative mechanism such as
the flux tube model.

The critical electron energy at which the dominant transport mechanism transitions from advection to ballistic can be estimated from the gyroradius of the electrons. For $1\,{\rm keV}$ synchrotron X-rays, they correspond to $30\,{\rm TeV}$ electrons assuming a magnetic field of $20\,{\rm\mu G}$, and the electron gyroradius would be $0.002\,{\rm pc}$. This is smaller than the observed $\sim0.1\,{\rm pc}$ radius of the X-ray tail so that these electrons are confined in the tail. MHD simulations have shown that a turbulent magnetic field can develop at scales of a few percent of the tail radius \citep{2019MNRAS.484.5755O}, corresponding to a few times $0.001\,{\rm pc}$. This is comparable with the gyroradius of these X-ray-emitting electrons, implying that they are not bound to these small-scale field lines and instead follow a ballistic transport along the large-scale magnetic field in the tail.

That being said, it is challenging to explain the gamma-ray observation with a
simple flux tube model. This is because, in our model, the particles continuously
leave the system and then become unobservable with a very low radiation efficiency in the flux tube, thus requiring a large injection
power from the pulsar. A simple leptonic fit to the TeV emission
yields a total electron energy of $2.6\times10^{45}$\,erg for an exponential
cutoff power-law distribution \citep{2025Innov...600802.}. Comparing this with
the particles' transversal time of $\sim10^9$\,s across a 10\,pc long tube
suggests that a power input of $\sim10^{36}$\,erg\,s$^{-1}$ is needed,
significantly larger than the spin-down luminosity of $2.3\times 10^{35}$ erg s$^{-1}$ for J1740.

The above estimate hints at additional emission from particles after they
left the flux tube. We model this as an additional diffusion zone. This is also
physically motivated: as the $B$-field inside the tube decreases to a strength
comparable to the ISM field, the two fields could mix such that particles are
scattered between field lines. This scattering randomizes the pitch angle
distribution, which makes the emission visible again.
Emission energy loss is significant when diffusion dominates as it increases
time for particles to travel through the region.

To model this, we follow the treatment by \citet{1995PhRvD..52.3265A} to derive
the particle evolution in a spherically symmetric diffusion zone. For a time
independent exponential cutoff power law distribution of injected particles,
i.e.\ $dN/d\gamma dt=Q_0\gamma^{-p}\exp(-\gamma/\gamma_{\rm cutoff})$, where
$Q_0$ is the normalization factor to be determined by matching the observed flux
and $\gamma$ the Lorentz factor of the electrons with a high energy cutoff at
$\gamma_{\rm cutoff}$, the resulting distribution function of electrons as a
function of radius $r$ and age $t$ is
\begin{equation}
    f(r,t,\gamma)=\frac{Q_0\gamma^{-p}e^{-\gamma/\gamma_{\rm cutoff}}}{4\pi D(\gamma)r}{\rm erfc}\left(\frac{r}{2\sqrt{D(\gamma)t_\gamma}}\right).
\end{equation}
Here $t_\gamma=\min(t,0.75/(p_2\gamma))$ is an empirical estimate of whether the
transport of particles at given energy $\gamma$ is limited by the age or the
cooling timescale. We assume a quadratic cooling relation of
$dE/dt=p_2\gamma^2$ with the constant $p_2$ that accounts for both IC and
synchrotron processes. The diffusion coefficient is assumed to follow a
power-law dependence of $D=D_{100}(E/100\,\mathrm{TeV})^\delta$, with
$\delta=1/3$ from the Kolmogorov turbulence model
\citep{1941DoSSR..30..301K,2017Sci...358..911A}.
We take $D_{100}=3.55\times10^{26}\,{\rm cm^2\,s^{-1}}$ based on the LHAASO
measurements \citep{2025Innov...600802.}. Finally, the emission from the diffusion zone is computed by
integrating the synchrotron and IC emission along the line of sight.

Note that we assume point injection and ignore the transition process from
anisotropic to isotropic pitch angle distributions when the particles enter the
diffusion zone \citep[see][]{2024NatAs...8.1284K}. We argue that this has no
effect on the X-ray emission morphology that we are interested in, since the
emission is mostly near the head of the magnetic flux tube. Depending on the
model parameters and observing energy band, the relative contribution between
the magnetic flux tube and the diffusion zone can vary significantly. A small
$\phi$ enhances the geometric boosting effect in the flux tube so that the
diffusion zone contribution is reduced. This also increases the magnetic field
near the PWN, therefore further enhancing the tail feature in X-ray. The
extended X-ray emission from the diffusion zone can be easily hidden by the
background, and the gamma-ray morphology will also deviate from the standard
Gaussian-like due to the contribution from the flux tube component.
In the limiting case of nearly face-on geometry, $\phi\approx0$, our model
would produce a gamma-ray point source surrounded by faint diffused emission.
This could explain some recently identified TeV halos that show two-component
structure with strongly suppressed diffusion
coefficient \citep{2017Sci...358..911A,2021PhRvD.104l3017R}.

\subsection{Modeling the X-ray and gamma-ray structure}

\begin{figure*}
\centering
\includegraphics[height=7cm]{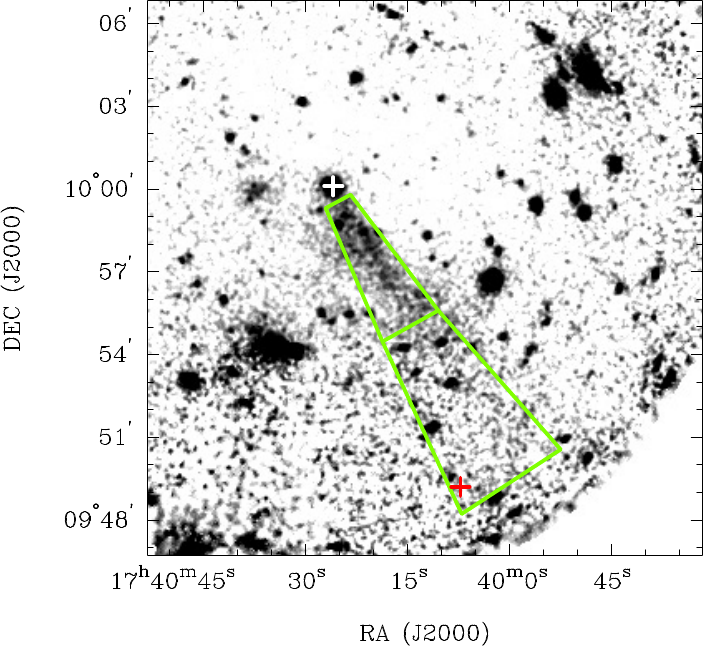}
\includegraphics[height=7cm]{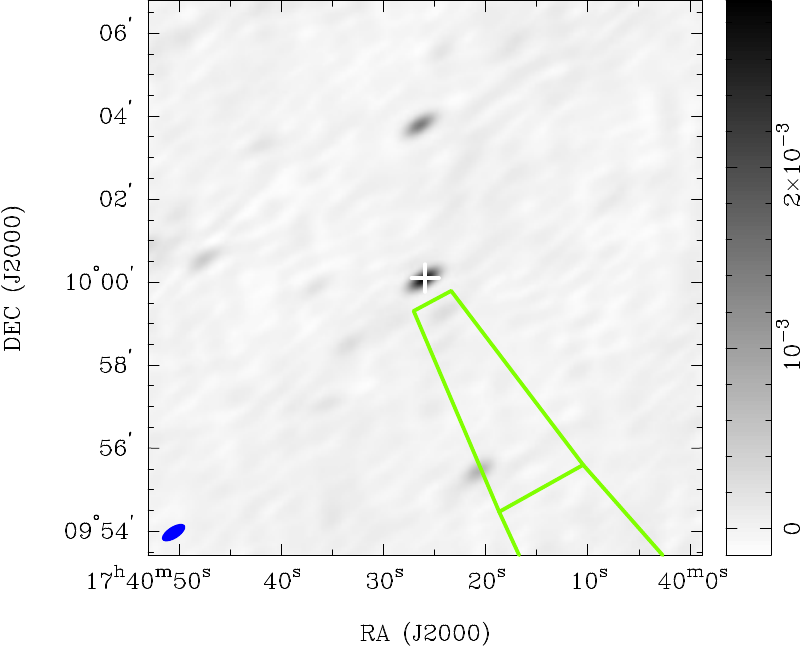}
\caption{Left: merged and exposure-corrected XMM-Newton image of J1740 tail 
from MOS 1 and MOS 2 in 0.4--7.2\,keV range (ObsIDs: 0803080201, 0803080301,
0803080401, and 0803080501). The white cross marks the position of J1740
and the red cross indicates the center of the TeV source 1LHAASO J1740+0948u.
The green region shows the tail as detected by XMM and the full extension assumed in our modeling. Right: VLA total
intensity radio map of J1740 at 10\,cm. The beam size is shown in lower left.
The tail is not detected in the data.
\label{fig:j1740}}
\end{figure*}

\begin{figure*}
\includegraphics[width=0.48\textwidth]{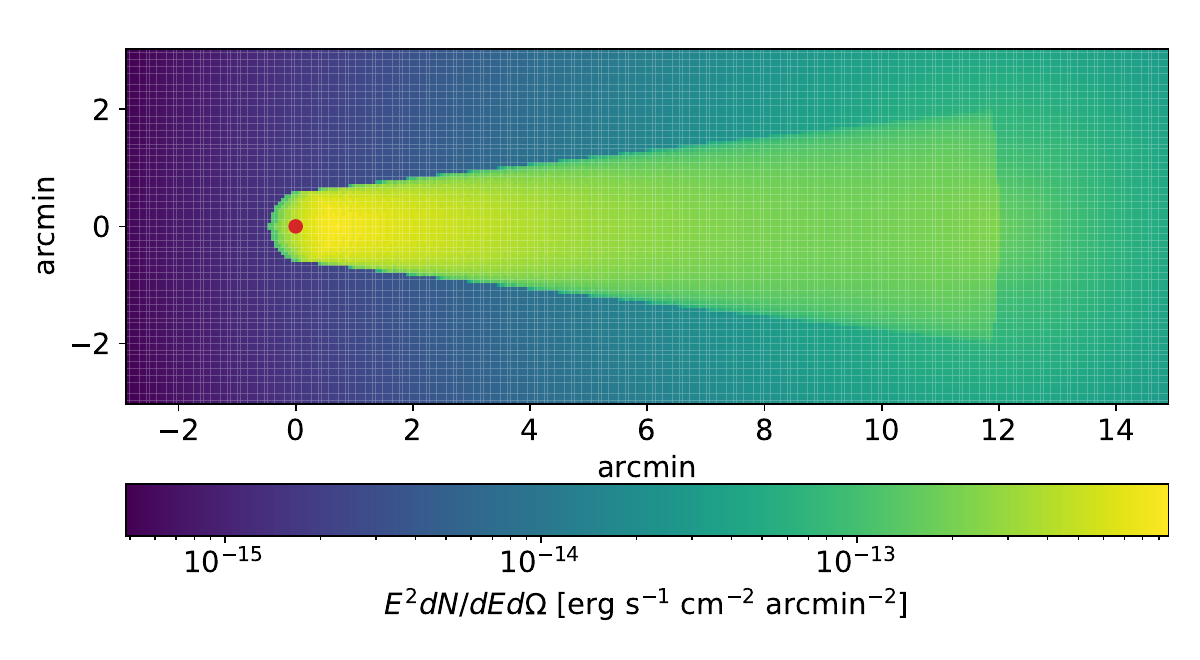}
\includegraphics[width=0.48\textwidth]{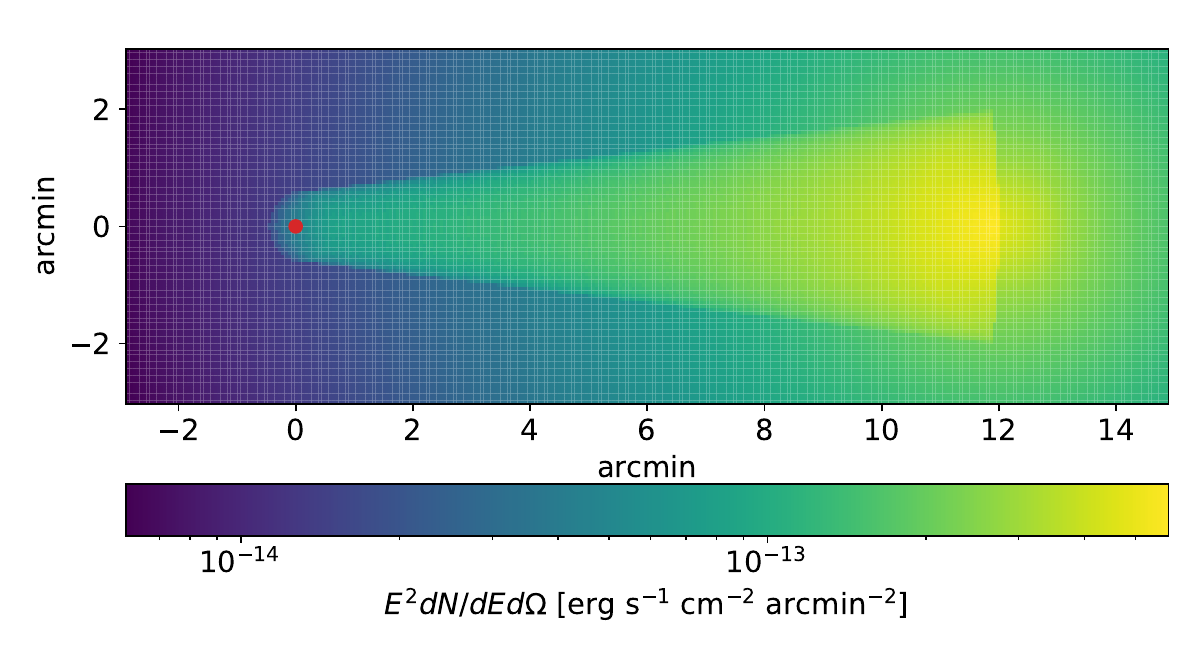}
\caption{Synchrotron X-ray image at 1\,keV (left) and IC gamma-ray image 10\,TeV
(right), generated using the best-fit parameters described in Section
\ref{sec:j1740}.
\label{fig:sim_j1740_image}}
\end{figure*}

In the modeling, we took $\phi=0.3$\,rad to match the opening angle of the J1740
tail. The length of the flux tube is set to $L=12'/\sin \phi$ such that
its projected length gives the 12\arcmin\ offset between the pulsar and the
gamma-ray center. We chose $R_L$ to match the observed 1.5\arcmin\ width of the
injection. The simulated X-ray and gamma-ray images are shown in
Figure~\ref{fig:sim_j1740_image}. Our model has successfully captured the
characteristics of the X-ray tail of J1740, i.e., the opening angle near the
pulsar and the bright emission in the first $\sim6'$. Beyond that, the
simulation predicts fainter emission extending all the way untill the end of the
12\arcmin\ tail with minor flux enhancement near the center of the diffusion
zone. The existing XMM data as shown in Figure \ref{fig:j1740} lack sensitivity to verify this.
Future observations with Einstein Probe Follow-up X-ray Telescope or eROSITA
will be useful. For gamma-ray emission, our model shows that it peaks near the
diffusion zone, which well agrees with the observation. We note that in
principle the flux tube can extend significantly beyond the cutoff point. In
this case, we expect two disjoint gamma-ray emission components, one near the
cutoff and one at the diffusion zone. These are however difficult to observe
given the poor angular resolution of gamma-ray observations in general. If a
clear gap is found, it would imply a different $B$-field strength at the
diffusion zone than at the cutoff point of the flux tube.

\begin{figure}
\includegraphics[width=\columnwidth]{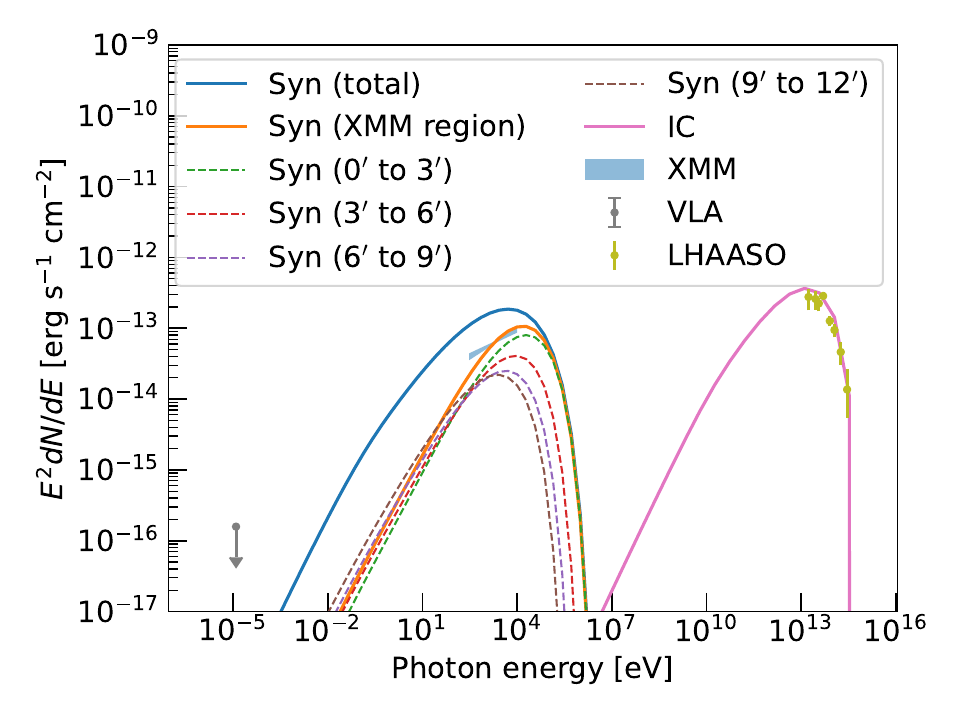}
\caption{Spectral fit of the two-component model to LHAASO \citep[yellow;][]{2025Innov...600802.}, XMM-Newton \citep[shaded blue;][]{2021ApJ...916..117B} and VLA (grey upper limit) data. The blue line shows the total synchrotron emission, while the orange line ignores weak extended emission and only considers a region of $1.5'\times6'$ so it can be compared with the XMM-Newton observation. The corresponding IC emission is shown as the purple line. The dashed lines show the emission in different sections along the flux tube direction, divided into segments of 3\arcmin\ length.
\label{fig:sim_j1740_spec}}
\end{figure}

\subsection{Modeling the broadband spectrum}

In addition to the emission structure, we also simulate the broadband spectrum
of J1740 based on the flux-tube-plus-diffusion-zone model above. We adopt the
exponential cutoff particle distribution with $E_{\rm cut}=110$\,TeV obtained
with LHAASO \citep{2025Innov...600802.}, and we also follow the same work to consider IC with cosmic
microwave background, near infrared (0.25\,eV\,cm$^{-3}$ at 30\,K), and
far infrared (0.5\,eV\,cm$^{-3}$ at 5000\,K) photons. The overall normalization
is set by the gamma-ray flux. This gives an injection rate of $3.23\times
10^{33}$\,erg\,s$^{-1}$, around $1.4\%$ of the spin-down rate of pulsar J1740.
Once the normalization is fixed, the $B$-field strength is determined by the
matching the X-ray flux. We consider only the region of $1.5'\times 6'$ near the
pulsar since the XMM measurement covers only a small field of view (see Figure \ref{fig:j1740}).
This suggests $B_L=1.2\,{\rm\mu G}$ at the exit end, corresponding to
$B_0=14\,{\rm\mu G}$ at the injection end. The former is consistent with the ISM
field strength at this high Galactic latitude \citep{2012ApJ...757...14J} and is
also compatible with the upper limit derived from the nondetection of diffuse
X-rays by XMM at the gamma-ray source location \citep{2025A&A...704A..30B}.
Figure~\ref{fig:sim_j1740_spec} shows the simulated broadband spectrum from
X-rays to gamma rays and a comparison with the data. We note that our model
gives a slightly harder X-ray spectrum ($\Gamma=1$) than the one observed by XMM
($\Gamma=1.7$), but closer to the Chandra measurement \citep[$\Gamma=1.2$;][]{2024ApJ...976....4D}. The softer spectrum can be explained if there is
additional emission component by particles not captured by the flux tube. We
also tried fitting with a softer injection spectrum and found that $p=2.5$ well
fits the X-ray spectrum but not the gamma-ray one. Finally in
Figure~\ref{fig:sim_j1740_spec} we plot the spatial variation of the synchrotron
X-ray spectrum along the flux tube. The spectral peak shifts to lower energy while
moving away from the pulsar. This is due to the decreasing magnetic field and
can be confirmed with spatially resolved spectroscopy from deep X-ray
observations in the future.

\section{Application 2: Misaligned PWN outflow}\label{sec:guitar}

Misaligned PWN outflows are found in several bow-shock PWNe in X-rays. They are
resulting from the injection of high-energy particles into well-ordered magnetic
filaments in the ISM. Our magnetic flux model therefore would be applicable
here. We focus on the most studied PWN filaments --- those in the Guitar Nebula
powered by the old pulsar B2224+65 with characteristic age $\tau_c$=1120\,kyr
and spin-down luminosity of $\dot E =1.2\times10^{33}$\,erg\,s$^{-1}$. The
pulsar has distance of 0.83\,kpc and proper motion of
$\mu=194.1\pm0.2$\,mas\,yr$^{-1}$ measured with parallax
\citep{2019ApJ...875..100D} Chandra X-ray observations show a 2.5\arcmin{} long
filament extending from the pulsar with a large (110\arcdeg) misalignment angle
from the proper motion direction
\citep{2022ApJ...939...70D,2024ApJ...976....4D}. Its X-ray spectrum is well
described by a power law with $\Gamma=1.6\pm1$ and an unabsorbed flux of
$6.35\times10^{-14}$\,erg\,s$^{-1}$\,cm$^{-2}$ in 0.5--7\,keV
\citep{2024ApJ...976....4D}. Detailed spatially resolved spectroscopy shows no
spectral softening along the filament except near the very end
\citep{2022ApJ...939...70D}, which is consistent with our flux tube model
prediction. The magnetic field inferred from equipartition is $\sim13\,\mu$G
near the pulsar and 8\,$\mu$G further away. A short counterfilament is also
observed, which extends at least 20\arcsec\ from the pulsar in a direction
opposite to the main filament.

In our modeling, we adopt a power-law particle distribution with index $p=2.2$
inferred from the X-ray spectrum. As the filament is only observed in X-rays,
we follow previous study to apply low- and high-energy cutoffs at 0.6 and
60\,TeV, respectively \citep{2008A&A...490L...3B}.
The observed filament does not strictly follow a truncated cone,
making it difficult to constrain the viewing angle. We nonetheless expect
the structure to lie mostly in the plane of the sky, given its 
nearly constant surface brightness.

\begin{figure}
\includegraphics[width=\columnwidth]{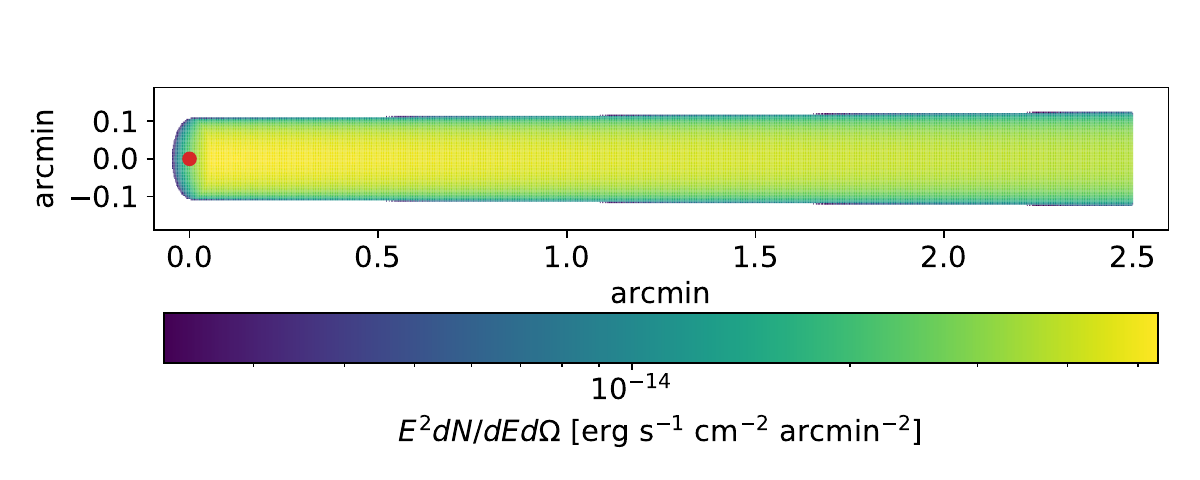}
\caption{Simulated magnetic flux tube image for the filament of Guitar Nebula at $1\,{\rm keV}$. The magnetic fields are assumed to be $56$ and $42\,{\rm\mu G}$ at two ends with a view angle of $\phi=\pi/3$. The projected length and width are chosen to reproduce the measured value.
\label{fig:sim_guitar_image_syn_1keV}}
\end{figure}

Figure~\ref{fig:sim_guitar_image_syn_1keV} shows an X-ray image based on the
flux tube model. This is generated with $\phi=\pi/3$ and $B$-fields of
56 and 42\,$\mu$G at the injection and exit ends, respectively.
The model captures the general features of the filament, and its finite length is
attributed to the cutoff in the pitch angle distribution.

The $B$-field required by our model is somewhat higher than the equipartition
values reported, but we note that the equipartition assumption may not hold as
particles traverse through the flux tube without much interaction with the
field. At the injection end the $B$-field is compatible with the typical PWN
field of the order of a few tens of $\mu$G. On the other hand, the exit end field
is larger than the typical ISM field. This can be reconciled if the flux tube
extends extend significantly beyond the cutoff point, so that the $B$-field continues.
Alternatively, plasma instability may boost the fields to such
\citep[e.g.,][]{2024A&A...684L...1O}.

Our calculation assumes 100\% of the pulsar spin-down power injected into
the flux tube. For a lower injection rate, a slightly higher magnetic field
strength is required to match the observed X-ray flux (synchrotron emissivity
scales as $B^2$). This can be relaxed if the injection process is time
dependent. Indeed, there is a hint of brightness fluctuation along the filament
between the year 2000 and 2006 epochs \citep{2022ApJ...939...70D}, which could imply
a variable injection rate. Finally, we are cautious that the flux tube model has
difficulty explaining the short counterfilament observed. The particles there
are likely moving away from us, and thus the synchrotron radiation beam would
become unobservable, unless there is an additional process, such as scattering,
that deflects the particles.

\section{Conclusion}

In this paper, we considered relativistic particles travel along a magnetic flux
tube with aligned field structure. The particles in this case would bound to
the field lines without much scattering. We show that in a decreasing
field configuration the particles' pitch angles will decrease, resulting in
a shrinking emission cone for the synchrotron and IC radiation.
This then creates a hard cutoff when the flux tube is viewed off axis.
We simulated X-ray and gamma-ray sky images of the flux tube model and found
that the synchrotron emission is brightest near the injection end where the
$B$-field is strong. This can create a linear tail-like feature in the X-ray
band. On the other hand, the IC gamma rays could show a significant offset from
the X-ray tail.

We applied the flux tube model to two PWN cases. First, we show that the X-ray
tail of J1740 and the associated TeV emission in the vicinity can be well
described by a flux tube plus a diffusion zone. Our model is able to produce the
morphology and spectrum of the broadband X-ray to gamma ray emission, and
provide constraints on the underlying particle distribution and magnetic field
strength. Second, we employed the flux tube
model to the misaligned filament of the Guitar Nebula. In our model, the
truncation of the filament is naturally explained by the cutoff due to viewing
geometry. The $B$-field we obtained for the Guitar Nebula is slightly higher than the typical PWN
value, but this can be reconciled if the flux tube extends beyond the cutoff
point.

In a future study, our modeling can be applied to other LHAASO sources associated
with bow-shock PWNe and pulsar filaments and, more generally, to any objects
with relativistic particles transported by a large-scale aligned magnetic field.

\begin{acknowledgments}
We thank the referee for the constructive advice. This work is supported by the National Natural Science Foundation of China under the grant No.~12375103.
This work is supported by a GRF grant of the Hong Kong Government under HKU 17301723.
The National Radio Astronomy Observatory and Green Bank Observatory are facilities of the U.S. National Science Foundation operated under cooperative agreement by Associated Universities, Inc.
Based on observations obtained with XMM-Newton, an ESA science mission
with instruments and contributions directly funded by
ESA Member States and NASA.
\end{acknowledgments}

\facilities{VLA, XMM}

\software{CASA \citep{2022PASP..134k4501C}, naima \citep{2015ICRC...34..922Z}, Miriad \citep{1995ASPC...77..433S}}

\appendix

\section{VLA observation of J1740}\label{sec:vla}

The VLA observation of J1740 was performed on 2015 October 23 at the 10\,cm band
in D configuration. The center frequency is at 3\,GHz with a total bandwidth of
2\,GHz and the total on-source time is 27 minutes. We processed the data
using the CASA package \citep{2022PASP..134k4501C}. We first flagged antennas
that missed calibrator observations and data affected by severe radio-frequency
interference. We then followed the standard procedures to calibrate the flux
scale, bandpass, and gains. The Stokes I image is formed using multifrequency
synthesis with a beam size of $0.64'\times0.29'$. The resulting image is shown
in Figure~\ref{fig:j1740}, which has an rms noise level of 41\,$\mu$Jy\,beam$^{-1}$. 
The pulsar is clearly detected with a flux density of $2.8\pm0.1$\,mJy, and the emission
is consistent with a point source. The PWN is not detected in the radio image.
For its size of $1.5'\times6'$ as found in the the XMM-Newton observation, we deduce
a $3\sigma$ flux density limit of 5.3\,mJy. This is plotted in Figure~\ref{fig:sim_j1740_spec} and compared with
the X-ray spectrum obtained with XMM-Newton. It is obvious that the limit is
incompatible with an extrapolation of the simple power-law X-ray spectrum.
These can be reconciled if there is a spectral break between the X-ray and radio bands.

\bibliography{export-bibtex}{}
\bibliographystyle{aasjournalv7}

\end{document}